\documentstyle[11pt,aaspp4]{article}
\lefthead{Brown, B. A. and Bregman, J. N.}
\righthead{X-Ray Emission from Early-Type Galaxies}

\begin{document}

\title{X-Ray Emission from Early-Type Galaxies: \\A Complete Sample Observed
by {\em ROSAT}} 

\author{Beth A. Brown and Joel N. Bregman} 
\affil{Department of Astronomy, University of Michigan, Ann Arbor, MI 48109-1090\\bab@umich.edu, jbregman@umich.edu}

\begin{abstract}
     To test the cooling flow model of early-type galaxies, we obtained a 
complete magnitude-limited sample of 34 early-type galaxies, observed with the 
PSPC and HRI on {\em ROSAT}.  The X-ray to optical distribution of galaxies 
implies a lower envelope that is consistent with the stellar emission inferred 
from Cen A.  When this stellar component is removed, the gaseous emission is 
related to the optical luminosity by $L_X \propto L_B^m$, where $m$ = 3.0-3.5, 
significantly steeper than the standard theory ($m$ = 1.7).  The dispersion 
about the correlation is large, with a full range of 30-100 in $L_X$ for fixed 
$L_B$.  The X-ray temperature is related to the velocity dispersion temperature 
as $T_X \propto T_{\sigma} ^n$, where $n$ = 1.43$\pm$0.21, although for several 
galaxies, $T_X$ is about twice $T_{\sigma}$.  The excessively hot galaxies are 
generally the most luminous and are associated with the richest environments.
     We suggest a model whereby environment influences the X-ray behavior of 
these galaxies:  early-type galaxies attempt to drive partial or total galactic 
winds, which can be stifled by the pressure of their environment.  Stifled winds should lead to hotter and higher luminosity systems, which would occur most 
commonly in the richest environments, as observed.
\end{abstract}


\section{Introduction} 

     The {\em Einstein Observatory} discovered that some early-type galaxies are powerful emitters of X-rays, and that their X-ray luminosity is correlated with 
their optical luminosity (Forman, Jones, \& Tucker \markcite{for85} 1985; 
Canizares, Fabbiano, \& Trinchieri \markcite{can87} 1987).  It also showed that the temperature of the hot gas is approximately that expected for gas that is 
gravitationally bound to the system.  For the more luminous X-ray-emitting 
ellipticals, where we are confident that the X-ray emission is dominated by the 
hot gas, the current model posits that gas is lost by stars through normal 
stellar evolution and is subsequently thermalized in the potential well of
the galaxy (e.g., Sarazin \markcite{sar90} 1990).  Radiative cooling is thought 
to exceed any heat sources (e.g., SNe), so that there is a net cooling, and the 
gas slowly flows into the center of the galaxy.

     The existing picture makes certain predictions, such as that the 
X-ray luminosity should be proportional to the mass loss 
rate of stars and the depth of the potential well (approximately $L_X \propto 
L_B^{1.7}$), with modest dispersion, and that the temperature of the hot gas should be proportional to the velocity dispersion 
squared.  Studies with the {\em Einstein Observatory} led to relationships that were either consistent with or steeper than the 
expected $L_X$-$L_B$ slope, but with considerable dispersion about the best-fit 
line (Canizares, Fabbiano, \& Trinchieri \markcite{can87} 1987; Donnelly, Faber \& O'Connell \markcite{don90} 1990; White \& 
Sarazin \markcite{whi91} 1991; Bregman et al.\ \markcite{bre92} 1992; 
Kim et al.\
\markcite{kim92} 1992).  The {\em Einstein Observatory} did not provide accurate temperature information for comparison with the velocity dispersion.  {\em ROSAT}, with a factor of two improvement in spectral resolution, does permit
accurate determinations of gas temperature.  The first published {\em ROSAT} sample is that of Davis and White \markcite{dav96} 
(1996), who found a correlation between X-ray temperature and stellar velocity 
dispersion, with temperatures greater than anticipated.  Here we present X-ray 
luminosities from our complete survey of optically selected 
elliptical galaxies, and temperatures for a subset of the sample.  An extensive discussion of the sample, data processing 
techniques, and additional statistical analysis of the sample will appear 
elsewhere (Brown \& Bregman 1998).

\section{Sample Selection and Observations}

     The primary goal was to define an unbiased sample of early type galaxies for which a complete set of X-ray data could be obtained.  We chose an 
optically-selected flux-limited sample of galaxies based upon the work of Faber 
et al.\ \markcite{fab89} (1989), who obtained a consistent set of distances for 
these galaxies.  This is essential for accurately establishing the $L_X$-$L_B$ 
relationship and defining the scatter about the relationship.

     The number of galaxies in the sample is a compromise between the desire to 
have many targets while avoiding many upper limits in the sample.  Based upon 
the $L_X$-$L_B$ relationship from the {\em Einstein Observatory}, we estimated 
that the percentage of detectable galaxies decreases significantly beyond the 
30--40 optically brightest galaxies, for sensible exposure times with the PSPC 
or HRI on {\em ROSAT}.  With 30--40 objects, our Monte-Carlo calculations showed that we should be able to define the $L_X$-$L_B$ relationship with sufficient 
accuracy to discriminate between competing models.  Our final sample (Table 1) is 
comprised of the 34 optically brightest early-type galaxies in Faber et al. 
\markcite{fab89} (1989) with $|b| > 20^{\circ}$, while avoiding dwarf galaxies 
(NGC 185, NGC 205, NGC 221) and X-ray bright quasars (e.g., M87).

     Most sample galaxies are well-known elliptical galaxies that were the 
primary targets of studies by other {\em ROSAT} observers, so the data products 
for these were obtained from the archive.  About one-third of the sample was 
observed for this project, completing the sample.  Most observations were obtained with the PSPC, but the last few sources were observed with the HRI.  
Every galaxy is detected, unlike earlier samples.  The event lists are screened 
for periods of high background, eliminating times for which the background count rate is greater than two times the modal value.  This leads to the removal of typically 5\% of the data (further details in Pildis et al. 
\markcite{pil95} 1995).  Point sources in the field of the galaxies were masked if their detection was $\geq 3\sigma$ above the surrounding regions. Flat-fielding corrections used the exposure maps provided by the SASS
processing. 

     It is unclear how to best choose the region within which to define the 
X-ray emission from ellipticals because the flux increases logarithmically with 
radius for a typical X-ray surface brightness distribution that decreases as 
$r^{-2}$ ($\beta$=0.5; at radii large compared to the core radius).  We have 
chosen to define the flux within an optically-defined radius since we are trying to test models related to the galaxy (e.g., mass loss from the stars).  We 
extract our signal within a radius of 4$r_e$ ($r_e$ is de Vaucouleurs 
half-light radius), within which 85\% of the optical light is contained.  The 
background is taken between 4--6.3$r_e$ for all galaxies, which removes the 
X-ray emission surrounding the galaxy, while maximizing the signal-to-noise.  
This method creates a very well defined $L_X$ for this sample, although we 
recoginize there may be an underestimation of the flux in some galaxies (discussed 
further in Brown \& Bregman 1998).  For a few galaxies, such as the weakest sources observed with the HRI, the signal is defined within 1$r_e$ 
and extrapolated to 4$r_e$ (for a $\beta$=0.5 model, the correction factor is 1.57) to maintain consistency within the sample.  

     Raymond-Smith thermal plasma models were fit to the spectra, under the 
assumption of a fixed cosmic abundance of 50\% (80\% for NGC 4472) and adopting 
the Galactic $N_H$ (fixed for each galaxy), which is accurate to about 5\% 
(Hartmann \& Burton \markcite{har97} 1997) in the northern hemisphere and to 
about 10\% in the southern hemisphere (Dickey \& Lockman \markcite{dic90} 1990).  In all cases, all PI channels were utilized except for NGC 3585, NGC 4621, and 
NGC 5102 where the softest channels were neglected.  For the 19 PSPC galaxies 
with enough counts to constrain the temperature (generally, $>$300 counts), a 
single-temperature model was fit.  If the resulting $\chi ^2$ was unacceptable, 
a two-temperature model was subsequently fit, with a ``hard''component whose 
temperature was fixed at 2 keV (presumably, reflecting the stellar binary 
contribution); the softer component is assumed to be from the hot 
gas.  There are no significant differences in the values determined using a 2 
keV component or a 5 keV component.

     Several of our galaxies have been observed by ASCA, so a comparison is 
worthwhile, although there are quantitative differences in the spectra.  The 
ASCA spectra are not extracted with the same background subtraction, partly due 
to limitations in the point spread function, so some cluster or group emission
may be included in the ASCA spectra.  Also, the low-energy cutoff in the ASCA 
response (near 0.5 keV) makes it difficult to fit spectra cooler than 0.6 keV, 
and there is some concern about the calibration at low energies (e.g., Sarazin, 
Wise, and Markevitch \markcite{sar98} 1998).  Nevertheless, for 
single-temperature spectral fits the derived temperatures from ASCA agree with 
the ROSAT results to typically 5-10\%. Two-temperature or multiple temperature 
fits are favored for nearly every early-type galaxy examined by Buote and Fabian \markcite{buo97} (1997), and in the cases where NH and the hot component are 
similar to our values (e.g., NGC 1404), they obtain the same temperature for the cooler, gaseous component.

\begin{deluxetable}{lrrrrrclrrrrr}
\tablecaption{Galaxy Properties. \label{tab1}}
\tablewidth{0pt}
\newcommand{\sm}{\footnotesize}
\tablehead{
\colhead{\sm Name} &
\colhead{\sm D} &
\colhead{\sm log$L_B$} &
\colhead{\sm log$L_X$} &
\colhead{\sm $T_{\sigma}$} &
\colhead{\sm $T_X$} &
\colhead{} &
\colhead{\sm Name} &
\colhead{\sm D} &
\colhead{\sm log$L_B$} &
\colhead{\sm log$L_X$} &
\colhead{\sm $T_{\sigma}$} &
\colhead{\sm $T_X$}\\
\colhead{} & \colhead{\sm (Mpc)} & \colhead{\sm ($L_{\odot}$)} & \colhead{\sm erg s$^{-1}$} & \multicolumn{2}{c}{\sm (keV)} &
\colhead{} &
\colhead{} & \colhead{\sm (Mpc)} & \colhead{\sm ($L_{\odot}$)} & \colhead{\sm erg s$^{-1}$} & \multicolumn{2}{c}{\sm (keV)}
}
\startdata
   {\sm N 0720} & {\sm 41.00} & {\sm 10.95} & {\sm 41.10} & {\sm 0.387} &  {\sm 0.436*}  &  & {\sm N 4278} & {\sm 29.40} &  {\sm 10.72} & {\sm 40.55} & {\sm 0.450 \phm{\tablenotemark{a}}} & {\sm \nodata\phm{*}} \nl
   {\sm N 1316} & {\sm 28.44} & {\sm 11.34} & {\sm 41.08} & {\sm 0.403} &  {\sm 0.351*}  &  & {\sm N 4365} & {\sm 26.66} &  {\sm 10.79} & {\sm 40.48} & {\sm 0.390 \phm{\tablenotemark{a}}} & {\sm 0.200\phm{*}} \nl
   {\sm N 1344} & {\sm 28.44} & {\sm 10.66} & {\sm 39.47} & {\sm 0.163} & {\sm \nodata\phm{*}} &  & {\sm N 4374} & {\sm 26.66} &  {\sm 10.99} & {\sm 41.09} & {\sm 0.524 \phm{\tablenotemark{a}}} & {\sm \nodata\phm{*}} \nl
   {\sm N 1395} & {\sm 39.80} & {\sm 11.02} & {\sm 41.04} & {\sm 0.424} &  {\sm 0.437*}  &  & {\sm N 4406} & {\sm 26.66} &  {\sm 11.10} & {\sm 41.80} & {\sm 0.397 \phm{\tablenotemark{a}}} & {\sm 0.823\phm{*}} \nl
   {\sm N 1399} & {\sm 28.44} & {\sm 10.88} & {\sm 41.44} & {\sm 0.610} &  {\sm 0.944\phm{*}}  &  & {\sm N 4472} & {\sm 26.66} &  {\sm 11.32} & {\sm 41.77} & {\sm 0.524 \phm{\tablenotemark{a}}} & {\sm 0.936\phm{*}} \nl
   {\sm N 1404} & {\sm 28.44} & {\sm 10.74} & {\sm 41.27} & {\sm 0.323} &  {\sm 0.557*}  &  & {\sm N 4494} & {\sm 13.90} &  {\sm 10.20} & {\sm 39.28} & {\sm 0.300 \tablenotemark{a}} & {\sm \nodata\phm{*}} \nl
   {\sm N 1407} & {\sm 39.80} & {\sm 11.16} & {\sm 41.34} & {\sm 0.517} &  {\sm 0.913\phm{*}}  &  & {\sm N 4552} & {\sm 26.66} &  {\sm 10.71} & {\sm 40.92} & {\sm 0.434 \phm{\tablenotemark{a}}} & {\sm 0.405*} \nl
   {\sm N 1549} & {\sm 24.26} & {\sm 10.73} & {\sm 40.04} & {\sm 0.267} &  {\sm 0.186\phm{*}}  &  & {\sm N 4621} & {\sm 26.66} &  {\sm 10.78} & {\sm 39.79} & {\sm 0.367 \phm{\tablenotemark{a}}} & {\sm \nodata\phm{*}} \nl
   {\sm N 2768} & {\sm 30.64} & {\sm 10.79} & {\sm 40.41} & {\sm 0.248} & {\sm \nodata\phm{*}} &  & {\sm N 4636} & {\sm 26.66} &  {\sm 10.96} & {\sm 41.81} & {\sm 0.232 \phm{\tablenotemark{a}}} & {\sm 0.717\phm{*}} \nl
   {\sm N 3115} & {\sm 20.42} & {\sm 10.83} & {\sm 39.74} & {\sm 0.450} & {\sm \nodata\phm{*}} &  & {\sm N 4649} & {\sm 26.66} &  {\sm 11.14} & {\sm 41.48} & {\sm 0.740 \phm{\tablenotemark{a}}} & {\sm 0.823\phm{*}} \nl
   {\sm N 3377} & {\sm 17.14} & {\sm 10.21} & {\sm 39.42} & {\sm 0.108} & {\sm \nodata\phm{*}} &  & {\sm N 4697} & {\sm 15.88} &  {\sm 10.58} & {\sm 40.13} & {\sm 0.173 \phm{\tablenotemark{a}}} & {\sm 0.206*} \nl
   {\sm N 3379} & {\sm 17.14} & {\sm 10.49} & {\sm 39.78} & {\sm 0.257} & {\sm \nodata\phm{*}} &  & {\sm N 5061} & {\sm 23.92} &  {\sm 10.53} & {\sm 39.54} & {\sm 0.233 \phm{\tablenotemark{a}}} & {\sm \nodata\phm{*}} \nl
   {\sm N 3557} & {\sm 47.98} & {\sm 11.10} & {\sm 40.61} & {\sm 0.541} & {\sm \nodata\phm{*}} &  & {\sm N 5102} &  {\sm 3.10} &   {\sm 8.95} & {\sm 37.70} & {\sm 0.500 \tablenotemark{a}} & {\sm \nodata\phm{*}} \nl
   {\sm N 3585} & {\sm 23.54} & {\sm 10.72} & {\sm 39.84} & {\sm 0.308} & {\sm \nodata\phm{*}} &  & {\sm N 5322} & {\sm 33.22} &  {\sm 10.80} & {\sm 40.11} & {\sm 0.319 \phm{\tablenotemark{a}}} & {\sm 0.205\phm{*}} \nl
   {\sm N 3607} & {\sm 39.82} & {\sm 11.18} & {\sm 40.82} & {\sm 0.390} &  {\sm 0.372\phm{*}}  &  & {\sm N 5846} & {\sm 46.72} &  {\sm 11.26} & {\sm 42.01} & {\sm 0.491 \phm{\tablenotemark{a}}} & {\sm 0.733\phm{*}} \nl
   {\sm N 3923} & {\sm 31.66} & {\sm 10.99} & {\sm 40.90} & {\sm 0.297} &  {\sm 0.549\phm{*}}  &  & {\sm I 1459} & {\sm 44.50} &  {\sm 11.14} & {\sm 41.19} & {\sm 0.601 \phm{\tablenotemark{a}}} & {\sm 0.414*} \nl
   {\sm N 4125} & {\sm 39.72} & {\sm 11.16} & {\sm 41.01} & {\sm 0.332} &  {\sm 0.283\phm{*}}  &  & {\sm N 7507} & {\sm 35.00} &  {\sm 10.82} & {\sm 40.13} & {\sm 0.361 \phm{\tablenotemark{a}}} & {\sm \nodata\phm{*}} \nl

\enddata

\tablenotetext{}{Distance, $L_B$, and $T_{\sigma}$ derived from Faber et al. (1989) values. $H_0$ = 50 km/s/Mpc used throughout.  Starred values in the $T_X$ column denote a two temperature fit value where the 2nd component is fixed at 2.0 keV.}
\tablenotetext{a}{Adopted value for $T_{\sigma}$.}

\end{deluxetable}

     For the extraction of the flux, we used either the fitted temperatures or, 
for objects with few counts, we usually assumed $T_X$ = 1.5$T_{\sigma}$.  For a 
few objects with very low values of 1.5$T_{\sigma}$, a minimum temperature of 
0.3 keV was assumed, since lower temperatures were clearly a poor fit, even for 
objects with only 100 counts.  We note that the luminosities are nearly 
independent of the metallicity assumed, and the temperatures have a modest 
dependence on the metallicity (for a metallicity that differs by a factor of 
two from our adopted value, the temperature changes by 10\% or less).

\section{Analysis and Interpretation of the Observations}

     In this analysis of the X-ray luminosities, we consider the picture by 
which the X-ray emission is due to a combination of hot gas and stellar X-ray 
sources.  The stellar X-ray contribution is expected to be proportional to the 
optical luminosity, while the hot gas component seems to increase more rapidly 
with the $L_B$ (e.g., $L_B^2$).  Consequently, at low optical luminosities, this
would lead to a minimum $L_X$, and there does appear to be a lower envelope to the 
$L_X$-$L_B$ distribution (seen first by Canizares, Fabbiano, \& Trinchieri
\markcite{can87} 1987), bounded by a line that is linearly proportional to $L_B$ (Fig.1).  This lower envelope can be compared to the estimate of the stellar 
component inferred from Cen A, the nearest non-dwarf elliptical galaxy.  It was 
not included in our sample because it has an AGN that contributes to the X-ray 
emission.  Cen A was studied by Turner et al.\ \markcite{tur97} (1997), who identify a hard component, attributed to the 
stellar sources, and a soft component ($kT$ = 0.29 keV), attributed to emission 
from diffuse hot gas.  These components have luminosities of 
1.7$\times$10$^{39}$ erg s$^{-1}$ (hard) and 1.8$\times$10$^{39}$ erg s$^{-1}$ 
(soft) in the 0.5-2.0 keV band (d=3.5 Mpc).  The optical luminosity is inferred 
from the spatial analysis of van den Bergh \markcite{van76} (1976), leading to a value of $L_B$ = 3.8$\times$10$^{10} L_{\odot}$.  The X-ray-to-optical luminosity 
ratio for the hard component corresponds to the lower envelope of the 
$L_X$-$L_B$ distribution, supporting the inference that it is due to stellar 
X-ray emission.  If one attributes both the soft and hard components to stellar 
emission in Cen A (e.g., Irwin and Sarazin \markcite{irw97} 1997), the resulting ``stellar'' line is higher in the $L_X$-$L_B$ plane by a factor of two, but 
still consistent with the bottom of the $L_X$-$L_B$ distribution.

\begin{figure}
\figurenum{1}
\plotone{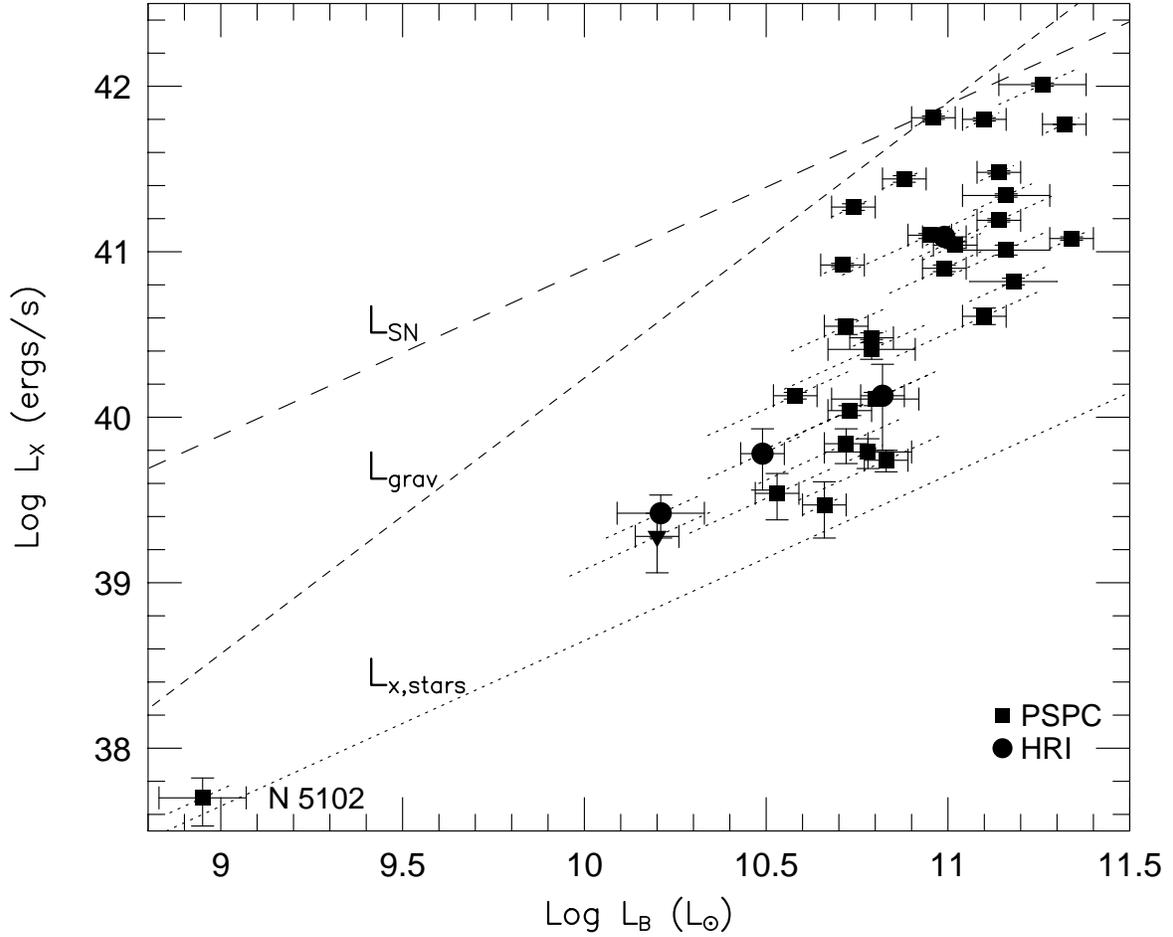}
\caption{The 0.5-2.0 keV luminosities from the {\em ROSAT} PSPC and HRI instruments are given 
against their optical blue luminosities.  The uncertainties due to distances are shown as dashed lines of slope unity while errors due to distance-independent 
effects (e.g., photon statistics) are shown as the usual horizontal and vertical lines.  The dotted line $L_{x,stars}$ is the stellar X-ray contribution as 
determined from Cen A, while the dashed lines labeled $L_{SN}$ and $L_{grav}$ 
represent the energy released from supernovae or available from thermalization 
and gravitational infall.}
\end{figure}

     The X-ray luminosity of the brightest galaxies can be compared to the maximum 
amount of energy 
expected from gravitational energy release and supernovae.  For the supernova 
rate, we use the mean values given by van den Bergh \& Tammann \markcite{van91} (1991) and a mean energy release per supernova of 10$^{51}$ ergs, which 
leads to log$L_{SN}$ = log$L_B$ + 30.89.  This supernova energy line bounds the 
distribution of X-ray emitting galaxies (Fig.\ 1), which indicates that supernova could provide sufficient energy to account for the hot X-ray emitting atmospheres of 
early-type galaxies.  We note that the rate of supernova energy input is 
probably uncertain by a factor of two or more due to the poor determination of 
the energy release per supernova and to the uncertainty in the supernova rate.

     The other main energy reservoir for the hot gas is the energy released as 
the mass loss from stars is thermalized and subsequently falls in the potential well 
of the galaxy.  If the gas falls nearly to the center before cooling, then for an 
isothermal atmosphere (we use fA=1, after Canizares, Fabbiano, \& Trinchieri 
\markcite{can87} 1987), and using the $L$-$\sigma$ relationship from Dressler 
\markcite{dre84} (1984) for Virgo ($L_B \propto \sigma^3$), we find that 
log$L_{grav}$ = 23.57 + (5/3)log$L_B$, which also lies along the upper bound of 
the $L_X$-$L_B$ distribution of the galaxies (Fig. 1).  Although there may be a 
factor of a few uncertainty in $L_{grav}$, this estimate shows that thermalization and gravitational release could be responsible for the observed $L_X$.

     The slope of the relationship between $L_X$ and $L_B$ is potentially quite 
important and has been the subject of much previous discussion.  If we exclude 
NGC 5102 (where the X-ray emission is consistent with a stellar origin), the log$L_X$-log$L_B$ relationship has a slope of 2.68$\pm$0.25, using the orthogonal linear regression bisector method (OLS bisector; Feigelson \& 
Babu \markcite{fei92} 1992); there is considerable scatter, approximately 1.5 
dex in $L_X$ for a fixed $L_B$.  When the hard, ``stellar'' component from Cen A is subtracted so 
that $L_X$ is only from the hot gas, the relationship becomes steeper, with a 
slope of 2.96$\pm$0.30, or 
3.51$\pm$0.41 (hard and soft components from Cen A removed).  This is much 
steeper than the slope of approximately 5/3 that is expected from the model 
where gas is thermalized, remains bound to the galaxy, and falls inward.

     The wide range of $L_X$ for a given $L_B$ and the steepness of the 
$L_X$-$L_B$ relationship for the gaseous emission are inconsistent with the
standard cooling flow model.  We suggest a modification of the model 
that may help to resolve these problems.  In this revised picture, environment 
plays a central role in determining the luminosity -- galaxies are trying to drive galaxtic winds, but the visibility of the X-ray emission depends on 
whether the wind is stifled (pressure confined) by the ambient cluster or group 
medium.  Observations support this in that the most X-ray luminous galaxies
are generally associated with the richest environments, such as the Virgo cluster (NGC 4406, NGC 4472, NGC 4636, NGC 4649), the Fornax cluster (NGC 1399, 
NGC 1404), or in the centers of moderate richness groups (e.g., NGC 5846).

     We find that many of these luminous galaxies are significantly hotter
than $T_{\sigma}$, another inconsistency with the basic model.  There is a 
correlation between the X-ray temperature, $T_X$, and the velocity dispersion 
temperature, $T_{\sigma}$, as is expected from theoretical arguments (Fig.\ 2), 
among the 19 galaxies for which temperatures were determined.  However, the 
slope of the log$T_X$-log$T_{\sigma}$ relationship is 1.43$\pm$0.21, which is 
steeper than the nominal theoretical expectation ($T_X$ = $T_{\sigma}$).  This 
is also significantly steeper than the slope of 0.73$\pm$0.10 reported by Davis 
\& White \markcite{dav96} (1996).  Also, Davis \& White never find 
$T_X$ = $T_{\sigma}$, which occurs for several of our galaxies.  We have 
determined that the difference between our slope with that of Davis \& White 
results partly from a difference in fitting procedures: we used the OLS bisector 
method, whereas Davis and White used y-on-x fitting (which yields a slope of 
0.90$\pm$0.28 for our data).  In addition, we find some galaxies systematically 
cooler than Davis \& White primarily because we allowed for a second hard component (a less important difference is that we fixed the metal abundance at 50\% solar 
rather than allowing it to be fit).

\begin{figure}
\figurenum{2}
\plotone{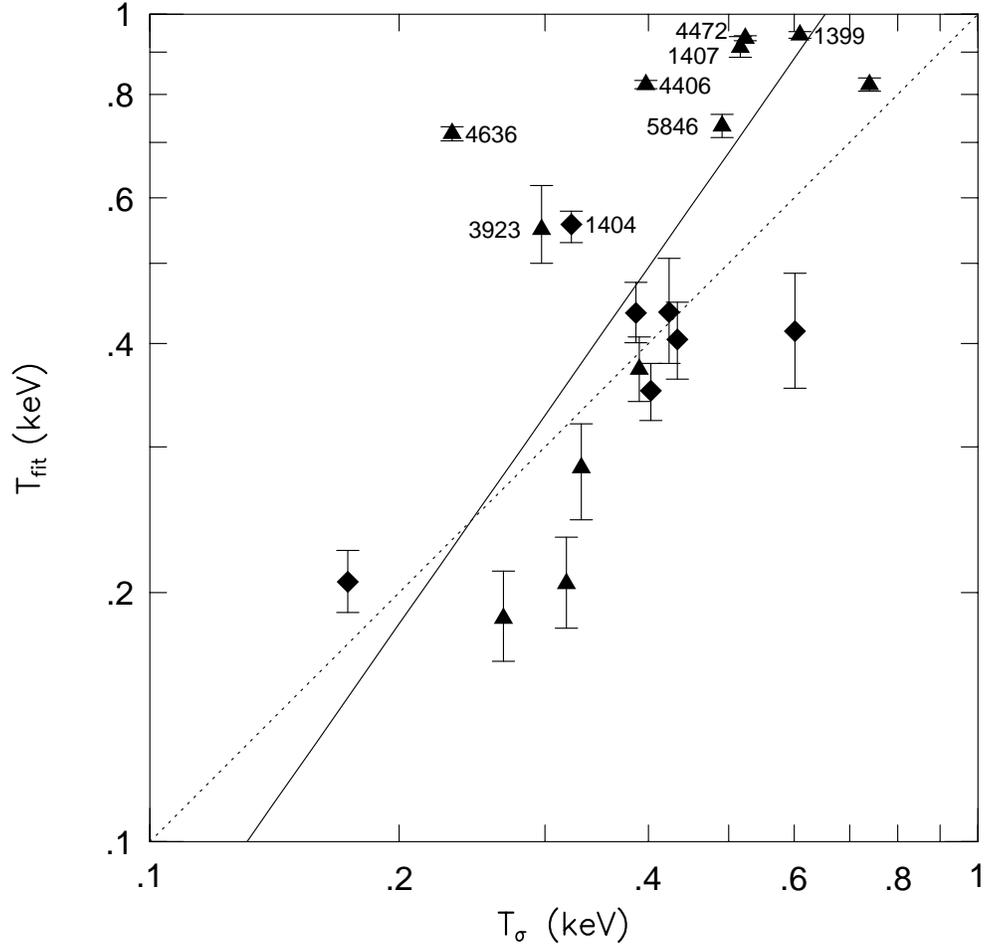}
\caption{The gas temperature, as determined from fitting the X-ray
spectral distribution vs the temperature corresponding to the velocity 
dispersion of the stars for high-count galaxies.  The triangles denote single-temperature fits while the diamonds denote two-temperature fits. The uncertainties shown are 90\% 
confidence errors.  The dashed line denotes the $T_X$ = $T_{\sigma}$ relation.
The solid line is the orthogonal linear regression bisector fit line.}
\end{figure}

     Two important features of the temperature distribution is that there is a 
substantial range in $T_X$ about either the best fit or the linear line, and 
that many of the galaxies have temperatures significantly above $T_{\sigma}$.  
Eight galaxies that are ``too hot'' relative to $T_{\sigma}$ are NGC 1399, NGC 
1404, NGC 1407, NGC 3923, NGC 4406, NGC 4472, NGC 4636, and NGC 5846. Six have 
X-ray temperatures that are about twice $T_{\sigma}$, and one galaxy, NGC 4636, 
is about three times hotter than its velocity dispersion temperature.  This list
is nearly identical, with an overlap of 7/8 galaxies, to a list of the eight
most X-ray luminous galaxies: NGC 1399, NGC 1404, NGC 1407, NGC 4406, 
NGC 4472, NGC 4636, NGC 4649, and NGC 5846 (in increasing NGC number).

     In our revised picture, the energy from supernovae is sufficient to heat the gas 
significantly above $T_{\sigma}$, thereby driving a partial galactic wind in the most 
massive galaxies and a more complete wind in less massive galaxies.  Galactic winds
that extend to very large distances transport most of the energy away as kinetic energy, so their X-ray luminosities can be very low, which would be consistent 
with the low values of $L_X$ for many of the galaxies in poor environments.  
Systems with partial galactic winds have a higher X-ray luminosity because the 
bound gas radiates its energy in the usual manner of a cooling flow; this would 
be common in the more massive galaxies.  However, if there is a high-pressure
ambient medium, as is found in a cluster, a terminal shock will occur near the 
galaxy, converting the kinetic energy to thermal energy, permitting the gas to 
accumulate, and creating a substantial emission measure (under some conditions, 
a galactic ``breeze'' solution may occur, which also should permit significant 
radiative losses).  The published supernova rates are sufficient to heat the gas
to 1 keV (or greater) and drive a wind (using either the rates of van den 
Bergh \& Tammann \markcite{van91} 1991 or Cappellaro et al. \markcite{cap97} 1997).
Also, the ambient pressure in environments such as the Virgo cluster (Nulsen \&
Bohringer \markcite{nul95} 1995) are adequate to balance the pressure in the
hot galactic medium (Brighenti \& Mathews \markcite{bri97} 1997) in the 10-100
kpc range.

     Environment may have several effects on the hot gas content of a galaxy.  
In a very dilute surrounding medium, a galactic wind would occur, but in a poor 
cluster, such as Virgo, the ambient medium may be adequate to pressure confine a wind, enhancing the X-ray luminosity of a galaxy.  A competing process is 
ram-pressure stripping, which would remove hot galactic gas,
ultimately reducing its X-ray luminosity (Takeda, Nulsen, \& Fabian 
\markcite{tak84} 1984; Gaetz, Salpeter, \& Shaviv \markcite{gae87} 1987; 
Sarazin \& White \markcite{sar88} 1988).  In very rich clusters, ram-pressure 
stripping should be the dominant process, so X-ray emission from hot 
galactic gas may be unimportant. 

     The primary weakness of our suggestion lies with the metallicity, which 
is expected to be above the solar value given the observed supernova rates
(Sarazin \markcite{sar97} 1997).  The X-ray metallicities would suggest that
the true supernova rate is 3-10 times lower than the published rates.  An 
alternative explanation is that the observed supernova rates are accurate but
that the metals are not mixed effectively into the hot galactic gas (see also
Ciotti et al. \markcite{cio91} 1991).  Ineffective
mixing of the metals would be necessary for our picture to remain viable.  
We look forward to the results of the ongoing supernova searches with 
CCD detectors, which should determine the observed rate to much higher accuracy.

\acknowledgements
     We would like to thank a variety of people for valuable
discussion: J. Irwin, J. Mohr, P. Hanlan, R. White,
M. Loewenstein, G. Worthey, J. Parriott, M. Roberts, D.
Hogg, and R. Mushotzky.  Special thanks is due to the members of
the {\em ROSAT} team and to the archiving efforts associated with the
mission.  Also, we wish to acknowledge the use of HEASARC and the NASA
Extragalactic Database (NED), operated by IPAC under contract with
NASA.  Support for this work has been provided by NASA through grants
NAGW-2135, NAG5-1955, and NAG5-3247; BAB would like to acknowledge
support through a NASA Graduate Student Researchers Program grant
NGT-51408.



\begin{references}

\reference{bau95}Bauer, F., \& Bregman, J. N. 1996, \apj, 457, 382
\reference{bre92}Bregman, J. N., Roberts, M. S., \& Hogg, D. E. 1992, \apj, 
  387, 484
\reference{bri97}Brighenti, F., \& Mathews, W.G. 1997, \apj, in press
\reference{buo97}Buote, D. A., \& Fabian, A. C. 1997, \mnras, in press
\reference{can87}Canizares, C. R., Fabbiano, G., \& Trinchieri, G. 1987, \apj, 
  312, 503
\reference{cap97}Cappellaro, E., Turatto, M., Tsvetkov, D.Y., Bartunov, O.S.,
  Pollas, C., Evans, R., \& Hamuy, M. 1997, \aap, 322, 431
\reference{cio91}Ciotti, L., D'Ercole, A., Pellegrini, S., \& Renzini, A. 1991,
  \apj, 376, 380
\reference{dav96}Davis, D. S., \& White, R. E. 1996, \apj, 470, L35
\reference{dic90}Dickey, J. M., \& Lockman, F. J. 1990, \araa, 28, 215
\reference{don90}Donnelly, R. H., Faber, S. M. \& O'Connell, R. M. 1990, \apj, 354, 52
\reference{dre84}Dressler, A., 1984, \apj, 281, 512
\reference{fab89}Faber, S. M., Wegner, G., Burstein, D., Davies, R. L., 
  Dressler, A., Lynden-Bell, D., \& Terlevich, R. J. 1989, \apjs, 69, 763
\reference{fei92}Feigelson, E. D., \& Babu, G. J. 1992, \apj, 397, 55
\reference{for85}Forman, W., Jones, C., \& Tucker, W. 1985, \apj, 293, 102
\reference{gae87}Gaetz, T. J., Salpeter, E. E., \& Shaviv, G. 1987, \apj, 316, 
  530
\reference{har97}Hartmann, D., \& Burton, W. B. 1997, ``Atlas of Galactic 
  Neutral Hydrogen'', Cambridge University Press
\reference{irw97}Irwin, J. A., \& Sarazin, C. L. 1997, \apj, in press
\reference{kim92}Kim, D. W., Fabbiano, G., \& Trinchieri, G. 1992, \apj, 393, 
  134
\reference{low91}Loewenstein, M., \& Mathews, W. G. 1991, \apj, 373, 445
\reference{low94}Loewenstein, M., Mushotzky, R. F., Tamura, T., Ikebe, Y., 
  Makishima, K., Matsushita, K, Asaki, H., \& Serlemitsos, P. J. 1994, \apj, 
  436, L75
\reference{nul95}Nulsen, P. E. J., \& Bohringer, H. 1995, \mnras, 274, 1093
\reference{pil95}Pildis, R. A., Bregman, J. N., \& Evard, A. E. 1995, \apj,
  443, 514
\reference{sar97}Sarazin, C. L. 1997, in The Nature of Elliptical Galaxies,
  ed. M. Amobili, G. DaCosta \&  P. Saha (San Francisco: PASP), in press
\reference{sar90}Sarazin, C. L. 1990, in The Interstellar Medium in Galaxies, 
  ed. H. A. Thronson, Jr. \& J. M. Shull (Dordrecht:  Kluwer), p. 201
\reference{sar88}Sarazin, C. L. \& White, R. E. 1988, \apj, 331, 102
\reference{sar98}Sarazin, C. L., Wise, M. W., and Markevitch, M. L. 1998, \apj, 
  in press.
\reference{tak84}Takeda, H., Nulsen, P. E. J., \& Fabian, A. C. 1984, \mnras, 
  208, 261
\reference{tur97}Turner, T. J., George, I. M., Mushotzky, R. F., \& Nandra, K. 
  1997, \apj, 475, 118
\reference{van76}van den Bergh, S. 1976, \apj, 208, 673
\reference{van91}van den Bergh, S., \& Tammann, G. A. 1991, \araa, 29, 363
\reference{whi91}White, R. E., \& Sarazin, C. L. 1991, \apj, 367, 476
\end{references}
\end{document}